\documentclass{PoS}

\usepackage{url}

\title{Performance of the Cherenkov Telescope Array}

\ShortTitle{Performance of the CTA}

\author{\speaker{G. Maier}$^{1}$, L. Arrabito$^{2}$, K. Bernl{\"o}hr$^{3}$, J. Bregeon$^{2}$, P. Cumani$^{4}$, T. Hassan$^{1}$, \mbox{J. Hinton$^{3}$}, \mbox{A. Moralejo$^{4}$} for the CTA Consortium\footnote{for consortium list see PoS(ICRC2019)1177} \\
       $^{1}$ Deutsches Elektronen-Synchrotron (DESY), Platanenallee 6, D-15738 Zeuthen, Germany\\
       $^{2}$Laboratoire Univers et Particules de Montpellier - UMR5299, Universit\'e de Montpellier - CNRS/IN2P3, Place Eug\`ene Bataillon - CC 72, 34095 Montpellier C\'edex 05 France\\
       $^{3}$Max-Planck-Institut f{\"u}r Kernphysik, P.O. Box 103980, 
        D-69029 Heidelberg, Germany\\
        $^{4}$Institut de Fisica d'Altes Energies (IFAE), The Barcelona Institute of Science and Technology, Campus UAB, 08193 Bellaterra (Barcelona) Spain\\
       E-mail: \email{gernot.maier@desy.de}}
       
\abstract{The Cherenkov Telescope Array (CTA) is expected to become the by far largest and most sensitive observatory for very-high-energy gamma rays in the energy range from 20 GeV to more than 300 TeV. 
CTA will be capable of detecting gamma rays from extremely faint sources with unprecedented precision on energy and direction. 
The performance of the future observatory derived from detailed Monte Carlo simulations is presented in this contribution for the two CTA sites located on the island of La Palma (Spain) and near Paranal (Chile). 
This includes the evaluation of CTA sensitivity over observations pointing towards different elevations and for operations at higher night-sky background light levels.}

\FullConference{36th International Cosmic Ray Conference -ICRC2019-\\
		July 24th - August 1st, 2019\\
		Madison, WI, U.S.A.}

\begin{document}

\section{Introduction}

The Cherenkov Telescope Array (CTA, \cite{CTA}) will be the most sensitive instrument for the observation of very-high-energy gamma rays,
providing a completely new view of the sky.
The observatory will be built at two sites in Paranal (Chile) and on La Palma (Spain), each consisting of a large number of imaging atmospheric Cherenkov telescopes.
CTA will observe the faint light induced through the Cherenkov effect by ultra-relativistic particles in the cascade initiated by high-energy gamma rays upon entering the atmosphere. 
The most important key performance improvements of CTA compared to currently operating instruments are:

\begin{itemize}

\item Two sites  in both hemispheres which provide a view of almost the entire sky.

\item A very large signal detection capability due to the employment of a large number of telescopes (99 telescopes at Paranal, 19 on La Palma). 
The effective area of CTA South is $5\times 10^4$ m$^2$ at 50 GeV, $10^6$ m$^2$ at 1 TeV, and beyond $5\times 10^6$ m$^2$  at 10 TeV.
This will provide orders of magnitude better sensitivity than e.g.~the Fermi LAT to short-timescale transient phenomena like GRBs or flaring active galactic nuclei \cite{Funk:2012}.

\item A powerful identification scheme of background events from  cosmic-ray nucleons, which results in an increase in sensitivity by a factor of five to ten as compared to the current instruments.

\item A very wide energy range from 20 GeV to beyond 300 TeV covered by a single facility through the deployment of telescopes with different optical collection areas.

\item A significant improvement in angular and spectral resolution. The angular resolution is expected to reach two arcminutes, allowing imaging of extended sources in great detail.
The improved energy reconstruction results in a resolution and systematic uncertainty on the energy scale of  less than 10\%.
This will provide the ability to observe new features in energy spectra (e.g.~lines or cutoffs) in moderately bright sources.

\item Flexibility in operations due to the large number of telescopes: feasible observation modes consisting of full array operation for highest sensitivity; sub-array operations for the simultaneous observations of several targets; and a divergent-pointing mode providing a instantaneous field of view of up to 20 deg diameter.

\end{itemize}

In this work, an overview of the CTA Observatory performance is provided and compared with the most important current and future observatories operating at these wavelengths.

\section{Monte Carlo simulations, reconstruction \& analysis}

Performance estimates are derived from detailed Monte Carlo simulations of the observatory  at the two CTA sites \cite{CTAMC}.
The simulations and analysis chain consists of air shower simulations and Cherenkov light production using the CORSIKA simulation code \cite{corsika}; the simulation of the detector response using the sim\_telarray package \cite{Simtelarray}, and reconstruction using the Eventdisplay \cite{Eventdisplay} and MARS \cite{MARS} analysis software packages\footnote{Dedicated CTA reconstruction software is currently under development and expected to provide improved performance compared to that presented here.}.

\begin{figure}
\centering\includegraphics[width=0.48\linewidth]{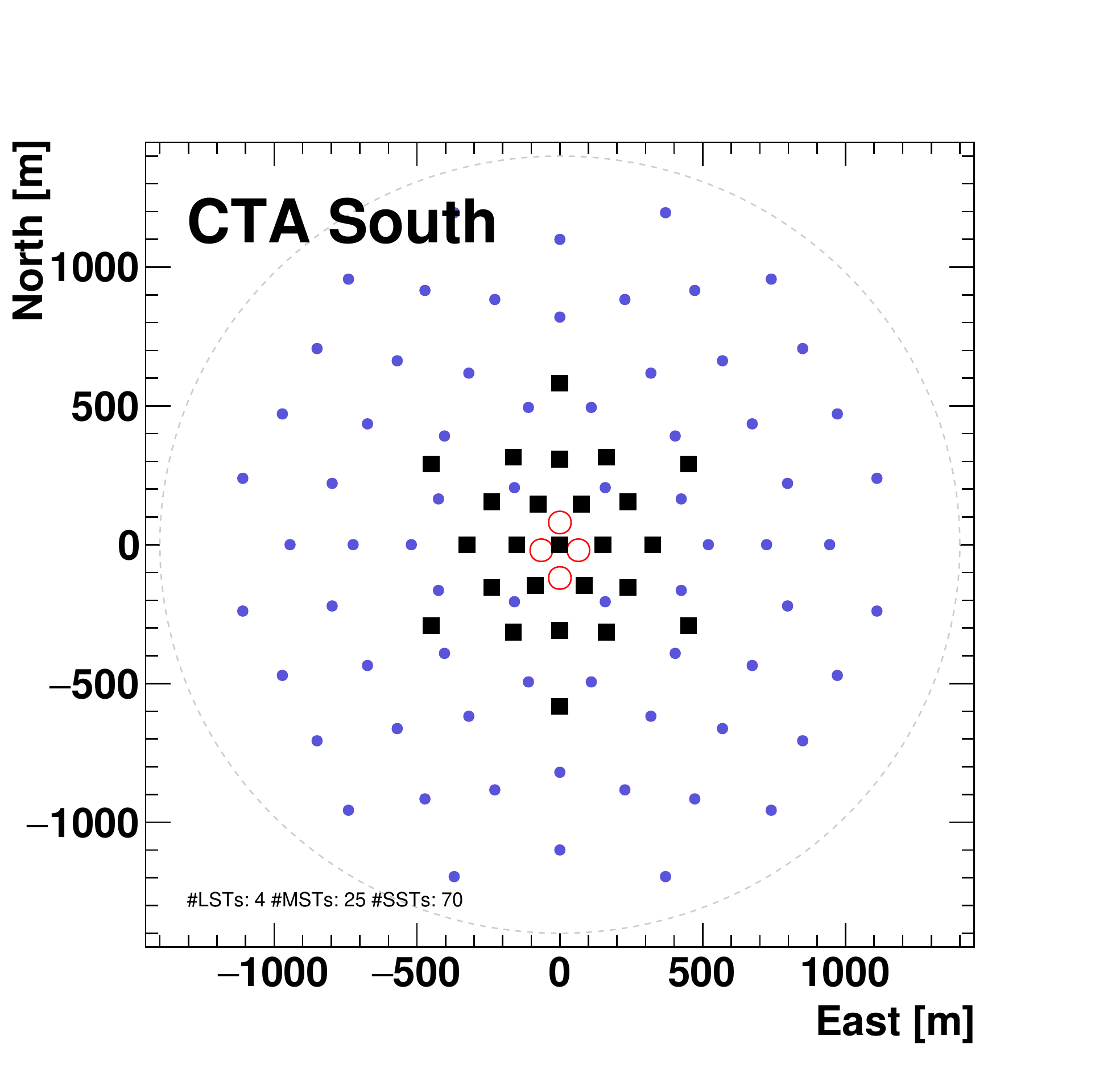}
\centering\includegraphics[width=0.48\linewidth]{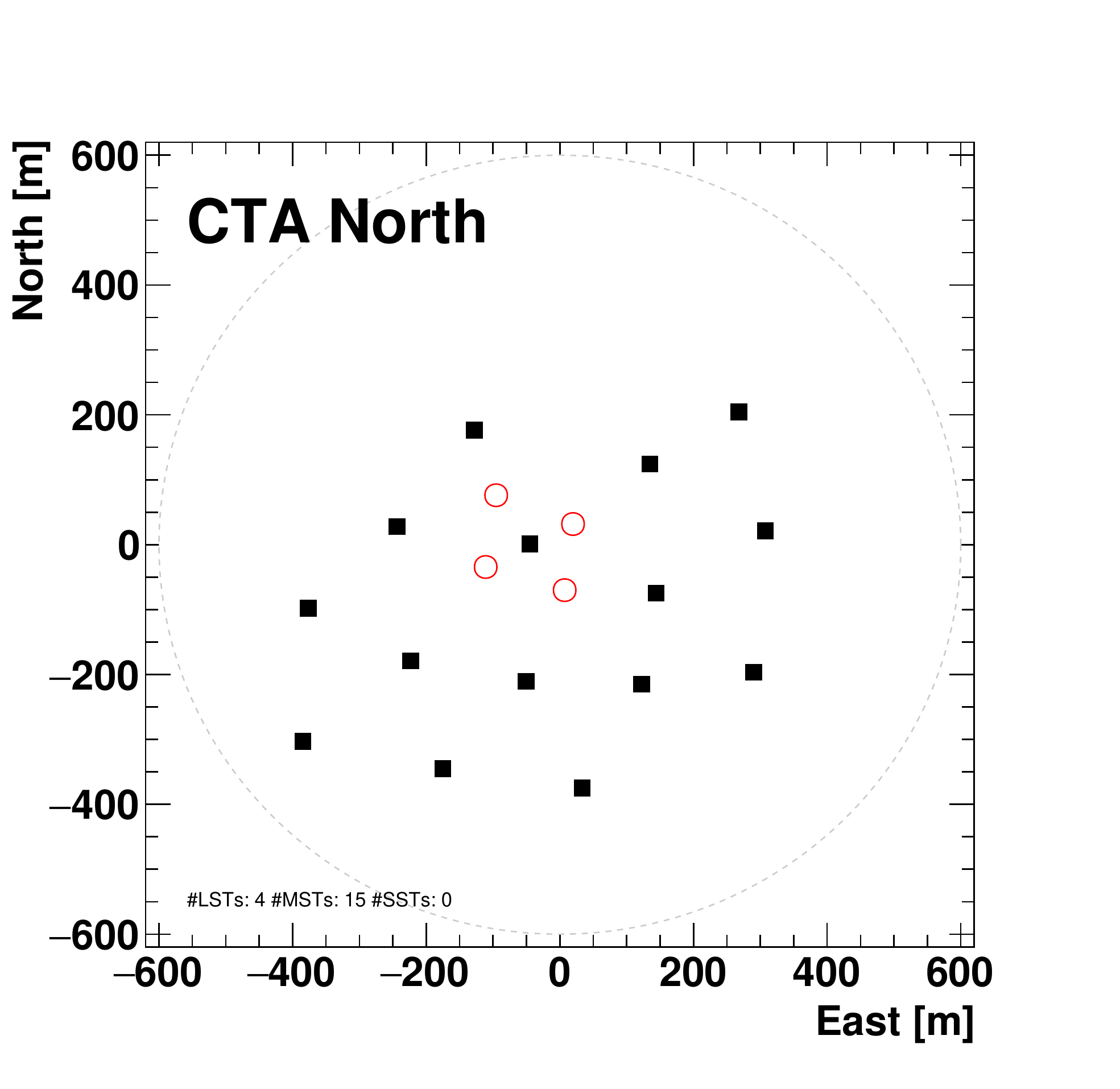}
\caption{\label{fig:array}
Telescope layouts for the Paranal (left) and La Palma site (right) of CTA.
Large-sized telescopes are indicated by open circles; medium-sized telescopes by filled squares; and small-sized telescopes by filled circles (southern site only).}
\end{figure}

Table \ref{tab:sites} gives an overview of the site characteristics and the number of telescopes of each type at the two CTA sites. 
Figure \ref{fig:array} shows the layout indicating telescope positions on the ground.
The site choice and the exact arrangement of the telescopes are the result of an extensive and detailed optimisation procedure \cite{CTAsites,CTAarrays}.
The Monte Carlo simulations take the local atmospheric conditions and the configuration of the geomagnetic field into account. 

\begin{table}[ht]
\centering
\begin{tabular}{c c c c  c c}
\hline
Site & Longitude, Latitude & Altitude  & LSTs & MSTs & SSTs   \\
 & [deg] & [m] &  \\
\hline
Paranal & 70.3W, 24.07S  & 2150  & 4 & 25 & 70  \\
\hline
La Palma & 17.89W, 28.76N & 2180  & 4 & 15 & -   \\
\end{tabular}
\caption{\label{tab:sites} Characteristics for the CTA Paranal and La Palma sites.
The number of telescopes of each type for each site are given for large-sized telescopes (LSTs), medium-sized telescopes (MSTs), and small-sized telescopes (SSTs).}
\end{table}

Background cosmic-ray spectra of proton and electron/positron particle types are 
set to match measurements from various cosmic-ray instruments. 
Heavier nuclei like cosmic ray helium are not simulated, as studies show that there is no significant contribution to the residual background after gamma-hadron separation cuts from these heavier nuclei.

The performance is evaluated for a point-like gamma-ray source located at the centre of the field of view of each camera (nominal telescope pointing scheme is assumed, with all telescopes pointing parallel to each other), with the exception of the results presented in Figure \ref{fig:diffsensOff}.

\section{Performance of CTA}

The unique capabilities of CTA for the detection of gamma rays is evaluated by the following metrics.

\begin{figure}
\centering\includegraphics[width=0.49\linewidth]{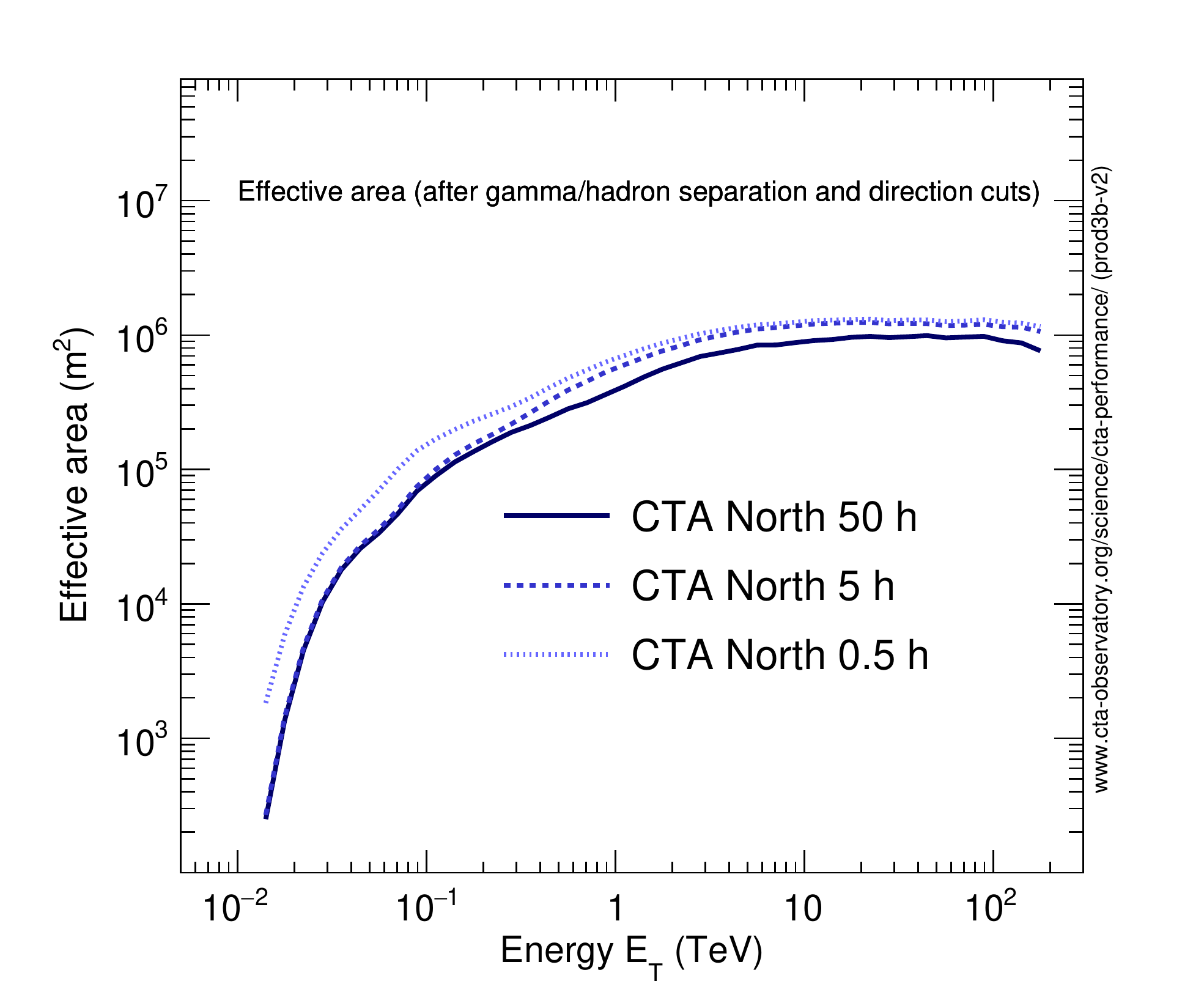}
\centering\includegraphics[width=0.49\linewidth]{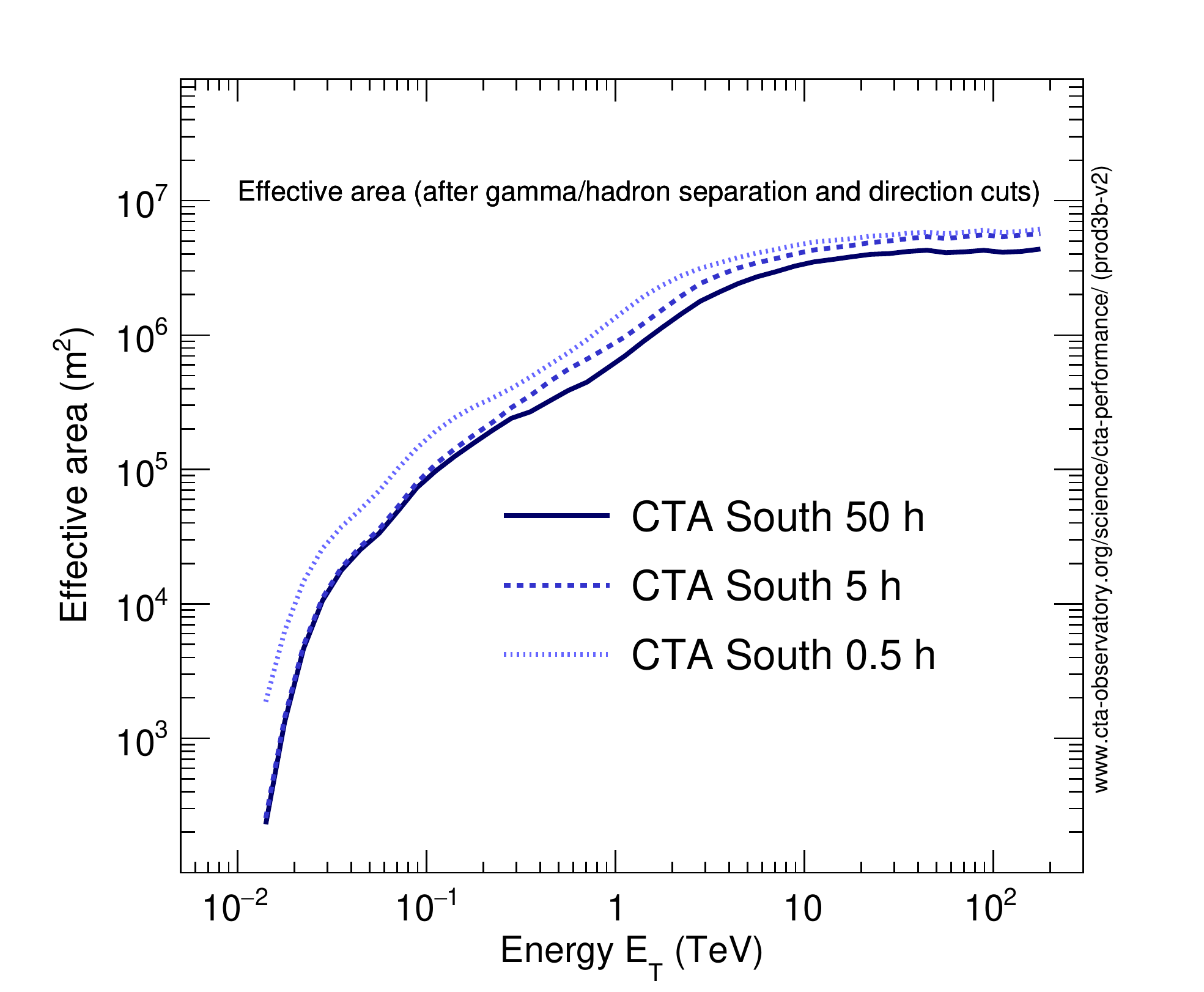}
\caption{\label{fig:effArea}
Effective collection area for point-like gamma-ray sources for CTA North (left) and CTA South (right).}
\end{figure}

The \textbf{effective collection area} for gamma rays describes the signal detection power of CTA.
The effective collection areas assuming point-like gamma-ray sources are shown in Figure \ref{fig:effArea} for cuts optimised to maximize sensitivity over a set of typical observation times.
It reaches $>10^6$ m$^2$ for CTA North and beyond $5\times 10^6$ m$^2$ for CTA South at high energies.
Especially notable is the sensitive area of several thousands of m$^2$ in the threshold region around 30 GeV.

\begin{figure}[ht!]
\centering\includegraphics[width=0.49\linewidth]{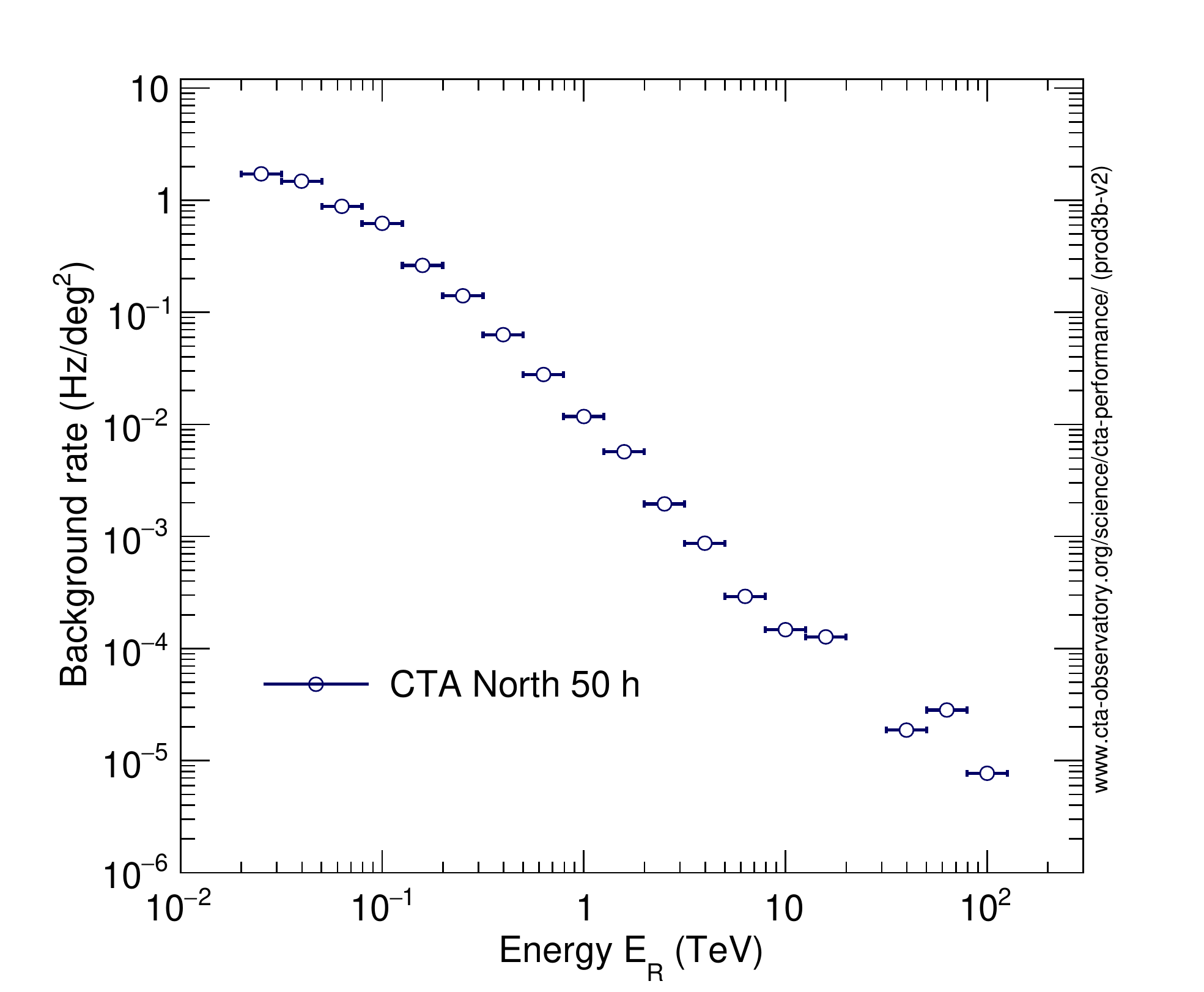}
\centering\includegraphics[width=0.49\linewidth]{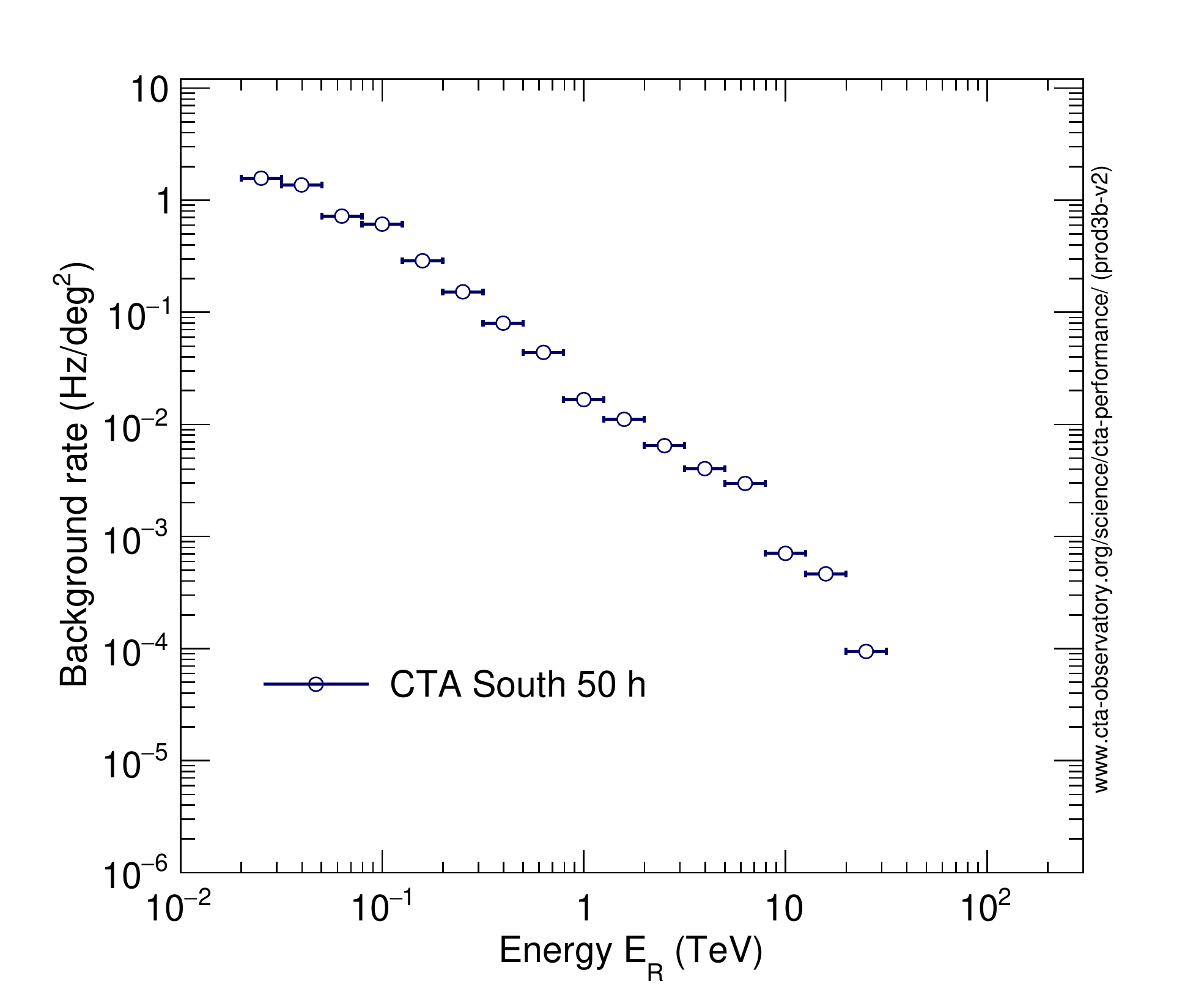}
\caption{\label{fig:background}
Residual background cosmic-ray background rates for gamma-hadron separation cuts optimised for 50 h of observation time.}
\end{figure}

The post-analysis \textbf{residual cosmic-ray background rate} for gamma-hadron separation cuts optimised for 50 h of observation time are shown in Figure \ref{fig:background}.
The background rate is integrated in 0.2-decade-wide bins in estimated energy (i.e.~five bins per decade). 
Note that the strong background suppression capabilities of CTA means that the majority of background events in the energy range between 200 GeV and $\approx 1.5$ TeV are due to cosmic-ray electrons and positrons.

\begin{figure}[ht!]
\centering\includegraphics[width=0.49\linewidth]{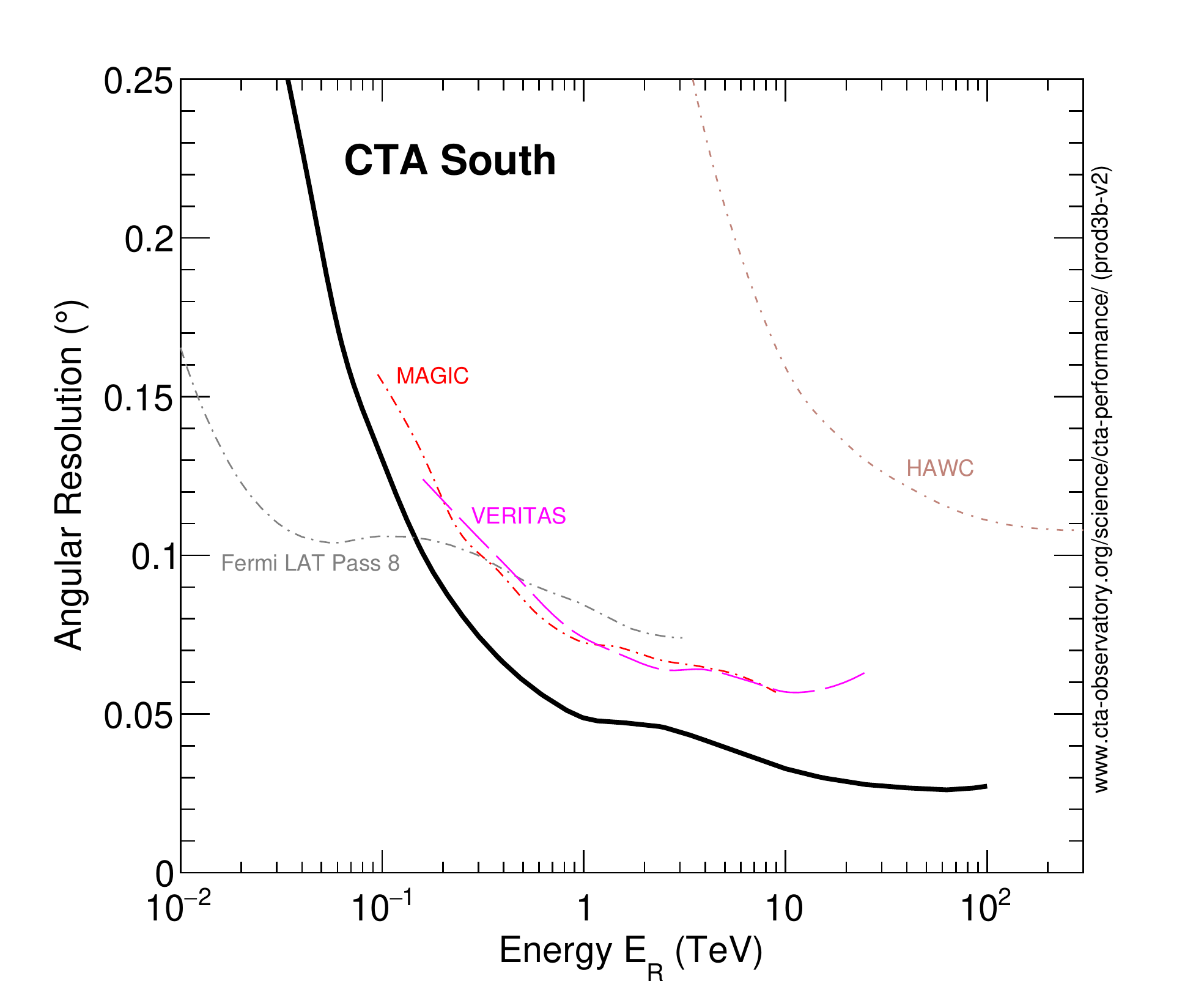}
\centering\includegraphics[width=0.49\linewidth]{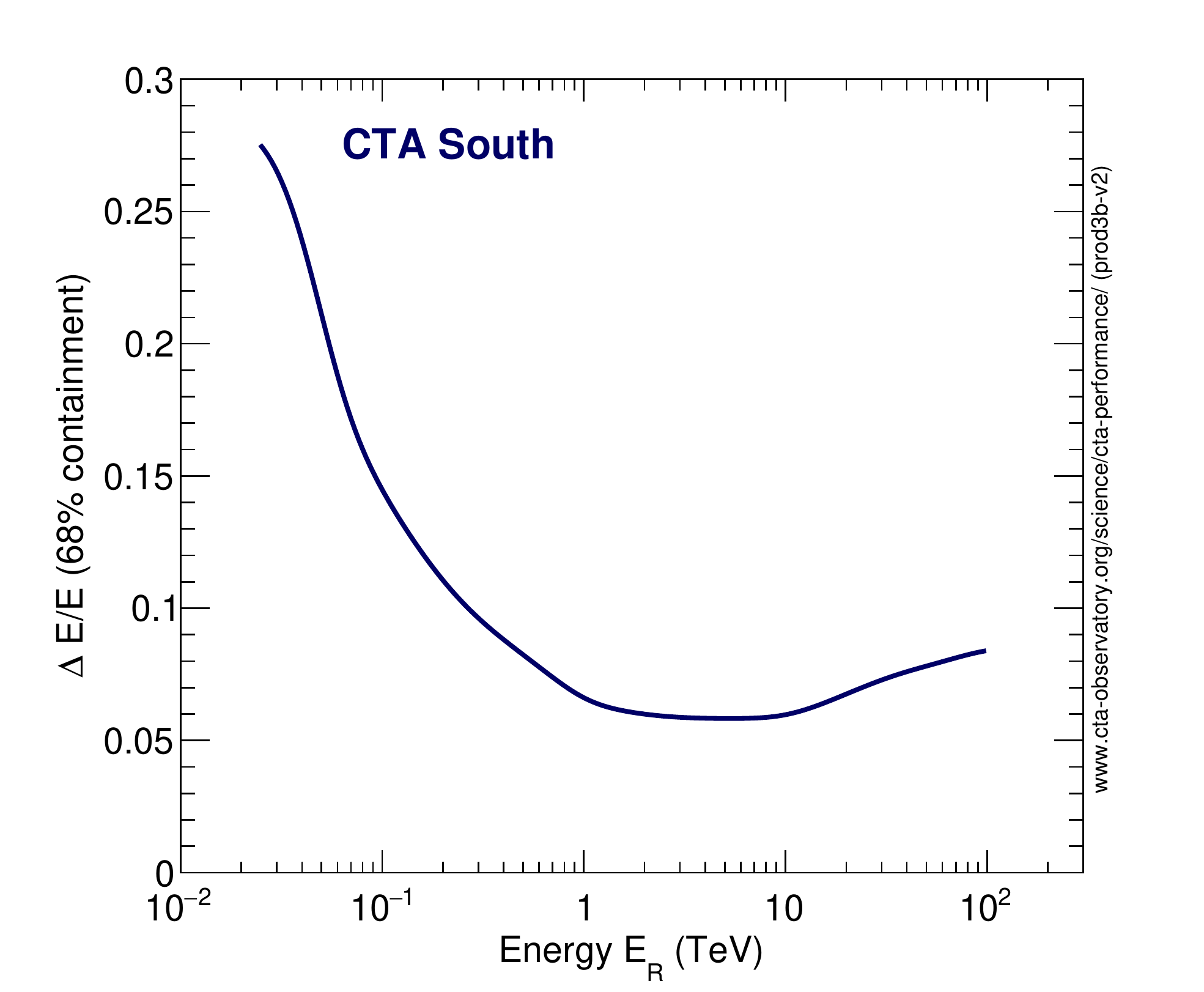}
\caption{\label{fig:resolution}
Left: Angular resolution vs reconstructed energy for CTA in comparison to existing gamma-ray instruments.
Right: Energy resolution vs reconstructed energy for CTA South.
Gamma-hadron separation cuts are applied for the MC events used to determine the angular and energy resolution.}
\end{figure}

The \textbf{angular resolution} is defined as the angle within which 68\% of reconstructed gamma rays fall, relative to their true direction (Figure \ref{fig:resolution}, left). 
CTA will achieve an angular resolution of better than 2 arcmin at energies above several TeV.
Note that this analysis is not optimised to provide the best possible angular resolution, but rather the best point-source sensitivity (as long as it complies with the minimum required angular resolution). 
Dedicated analysis cuts will provide, relative to the instrument response functions shown here, improved angular (or spectral) resolution, enabling e.g.~a better study of the morphology or spectral characteristics of bright sources.

The \textbf{energy resolution} $\Delta $E / E is obtained from the distribution of (E$_\mathrm{R}$ - E$_\mathrm{T}$) / E$_\mathrm{T}$, where E$_\mathrm{R}$ and E$_\mathrm{T}$ respectively refer to the reconstructed and true energies of gamma-ray events recorded by CTA (Figure \ref{fig:resolution}, right).
 $\Delta $E/E is the half-width of the interval around 0 which contains 68\% of the distribution.

\begin{figure}[ht!]
\centering\includegraphics[width=0.49\linewidth]{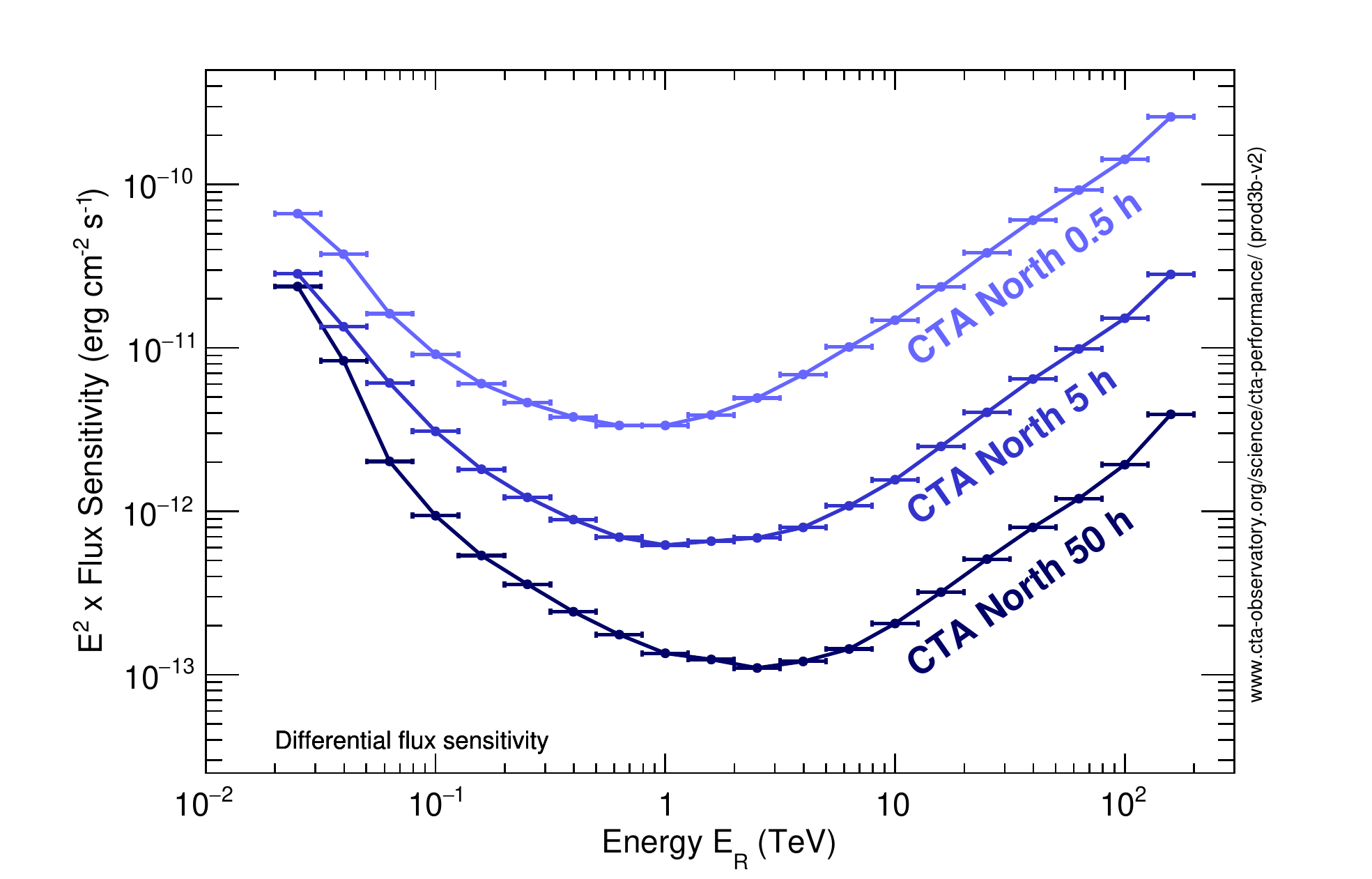}
\centering\includegraphics[width=0.49\linewidth]{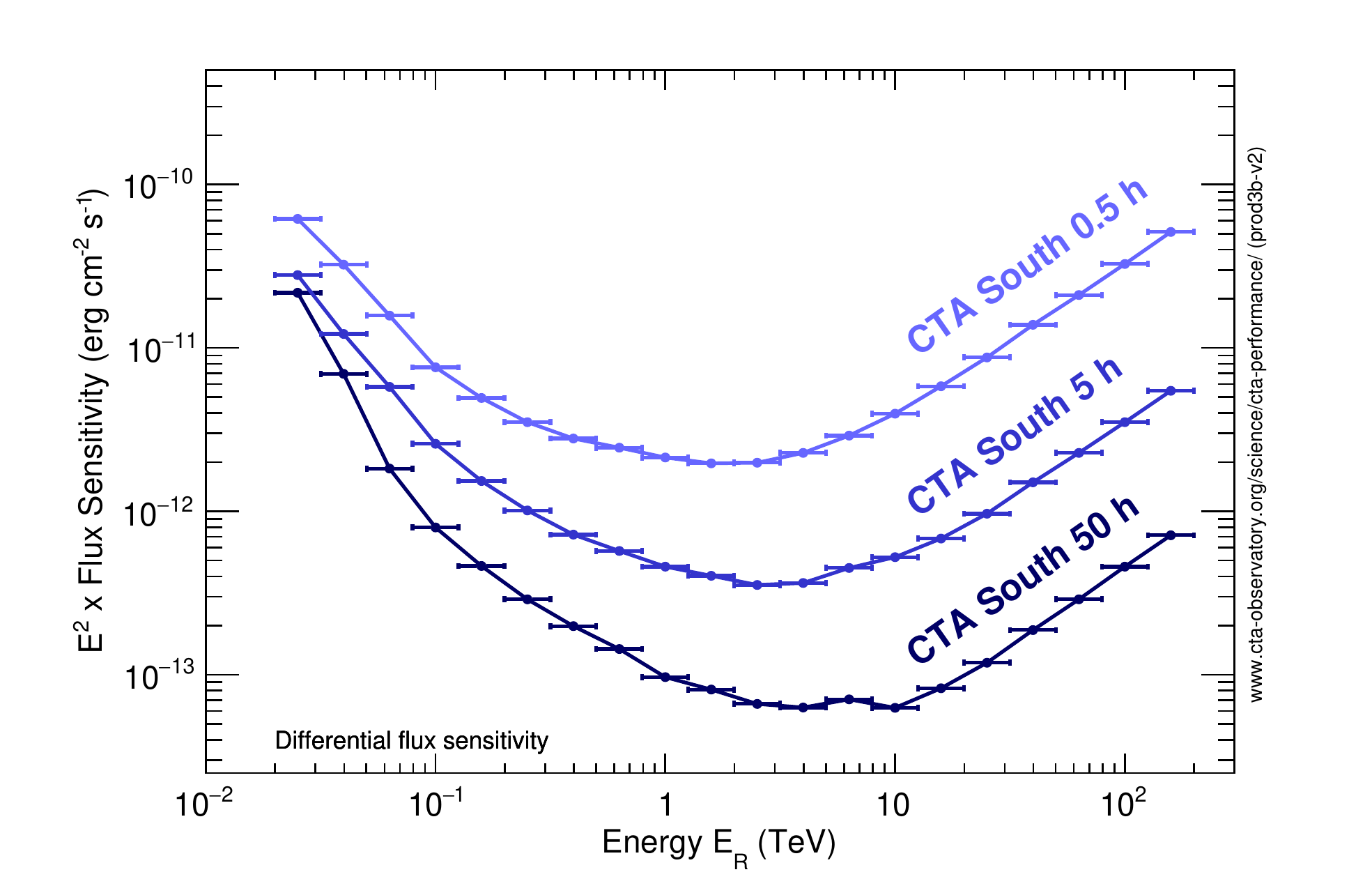}
\caption{\label{fig:diffsens}
Differential energy flux sensitivities for CTA North (La Palma site; left) and CTA South (Paranal site; right) for different observation times assuming a gamma-ray source located at the centre of the field of view.  
Detections are required in five independent logarithmic bins per decade in energy. 
Horizontal lines indicate the width of the energy bin.}
\end{figure}

%
\begin{figure}[ht!]
\centering\includegraphics[width=0.49\linewidth]{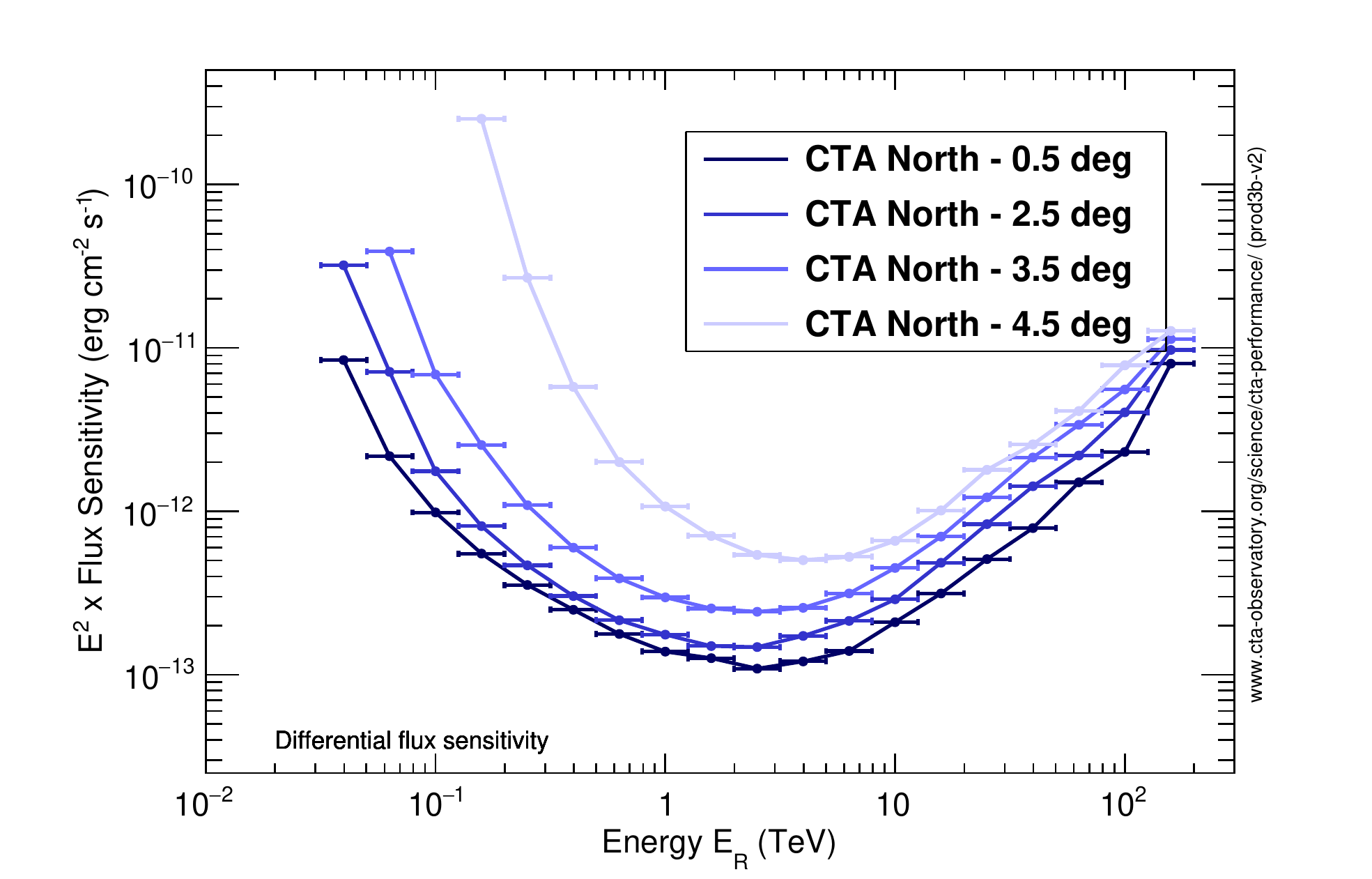}
\centering\includegraphics[width=0.49\linewidth]{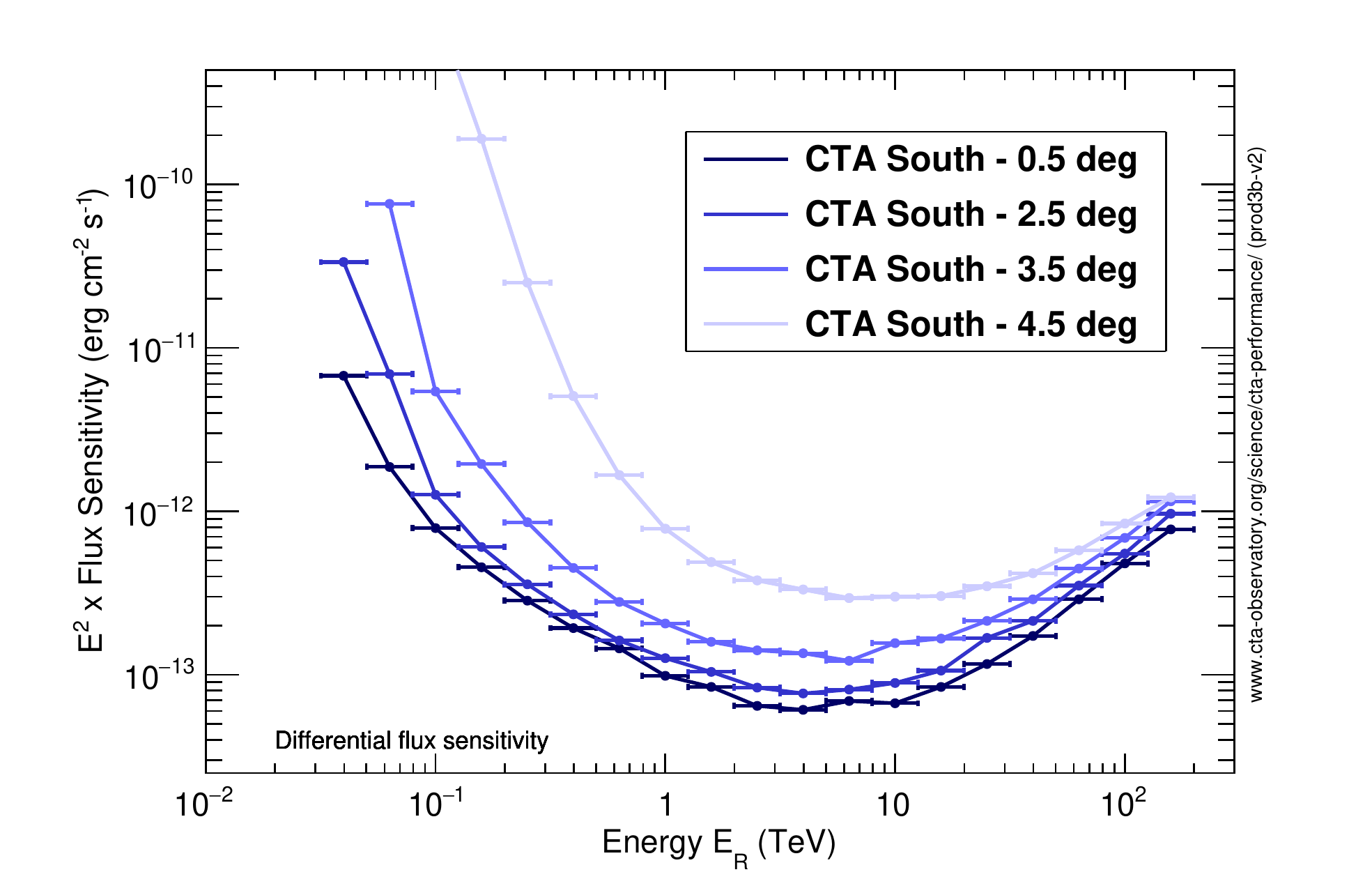}
\caption{\label{fig:diffsensOff}
Differential energy flux sensitivities for CTA North (La Palma site; left) and CTA South (Paranal site; right) for gamma-ray sources located at different offsets with respect to the centre of the field of view.  
An observation time of 50 h is assumed.
Detections are required in five independent logarithmic bins per decade in energy. 
Horizontal lines indicate the width of the energy bin.}
\end{figure}

\begin{figure}[ht!]
\centering\includegraphics[width=0.49\linewidth]{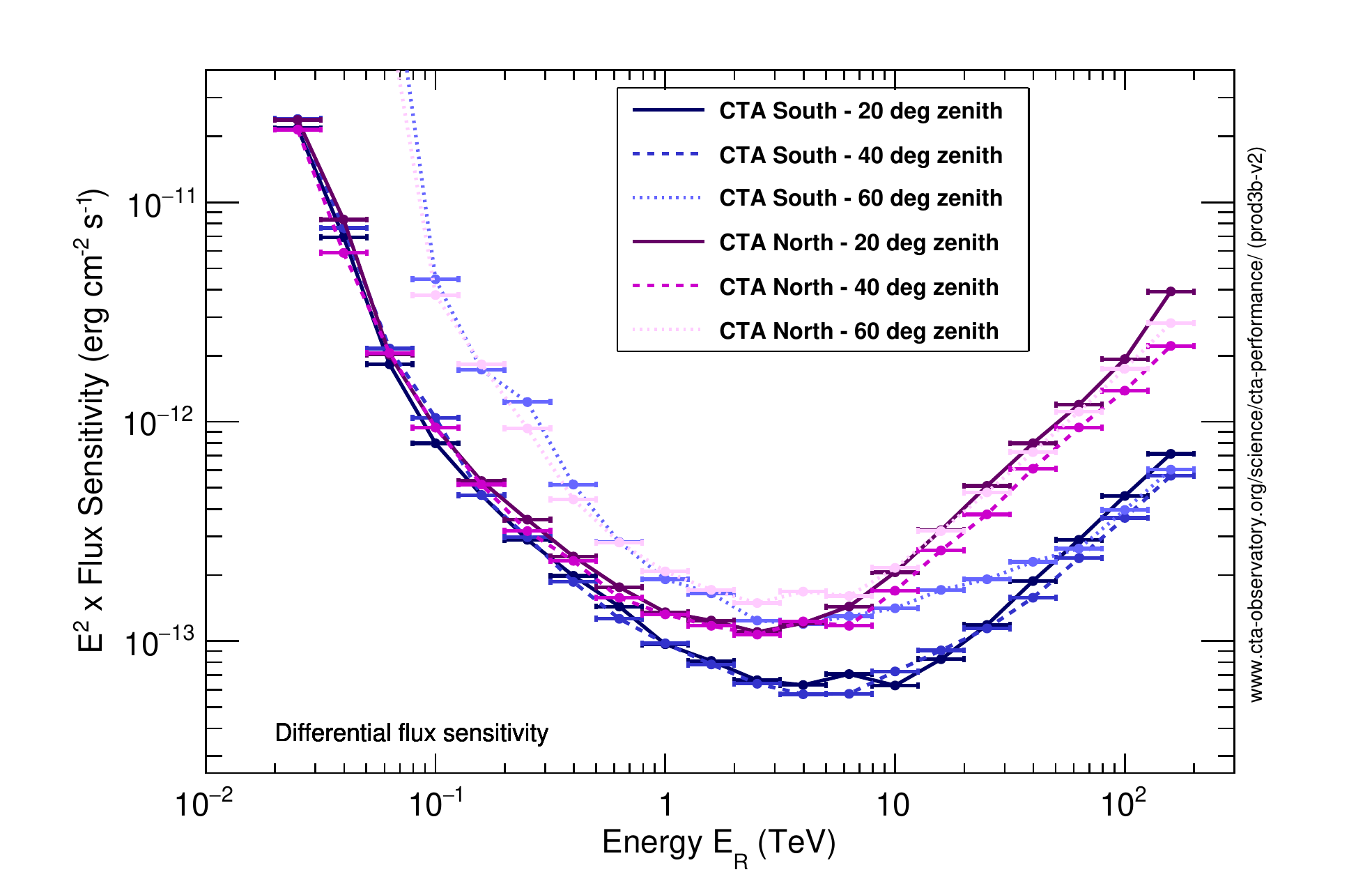}
\centering\includegraphics[width=0.49\linewidth]{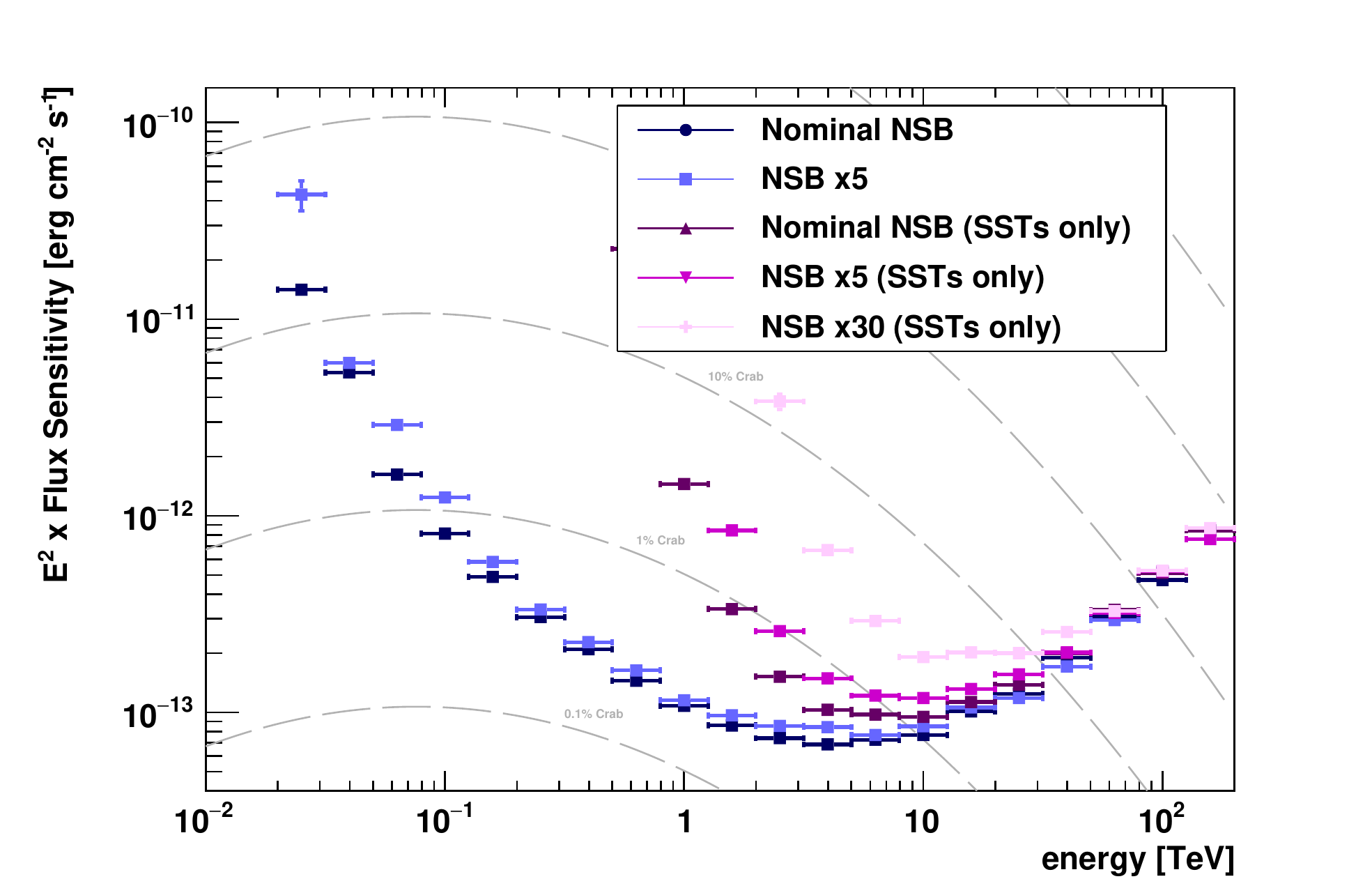}
\caption{\label{fig:diffsensNSB}
Left: Differential energy flux sensitivities for CTA North and South for observations at zenith angles of 20, 40, and 60 deg and gamma-ray sources located at  the centre of the field of view.
Right: Differential energy flux sensitivities for CTA South for observations at different levels of night-sky background light.
Curves for the full CTA array and subarrays of 70 small-sized telescopes only are shown.
The  background illumination level of $30\times$ the nominal dark environment refers to observations under very bright moonlight conditions and is planned for the SiPM-equipped small-sized telescopes only.
An observation time of 50 h is assumed. 
Detections are required in five independent logarithmic bins per decade in energy. 
Horizontal lines indicate the width of the energy bin.}
\end{figure}

\begin{figure}[ht!]
\centering\includegraphics[width=0.65\linewidth]{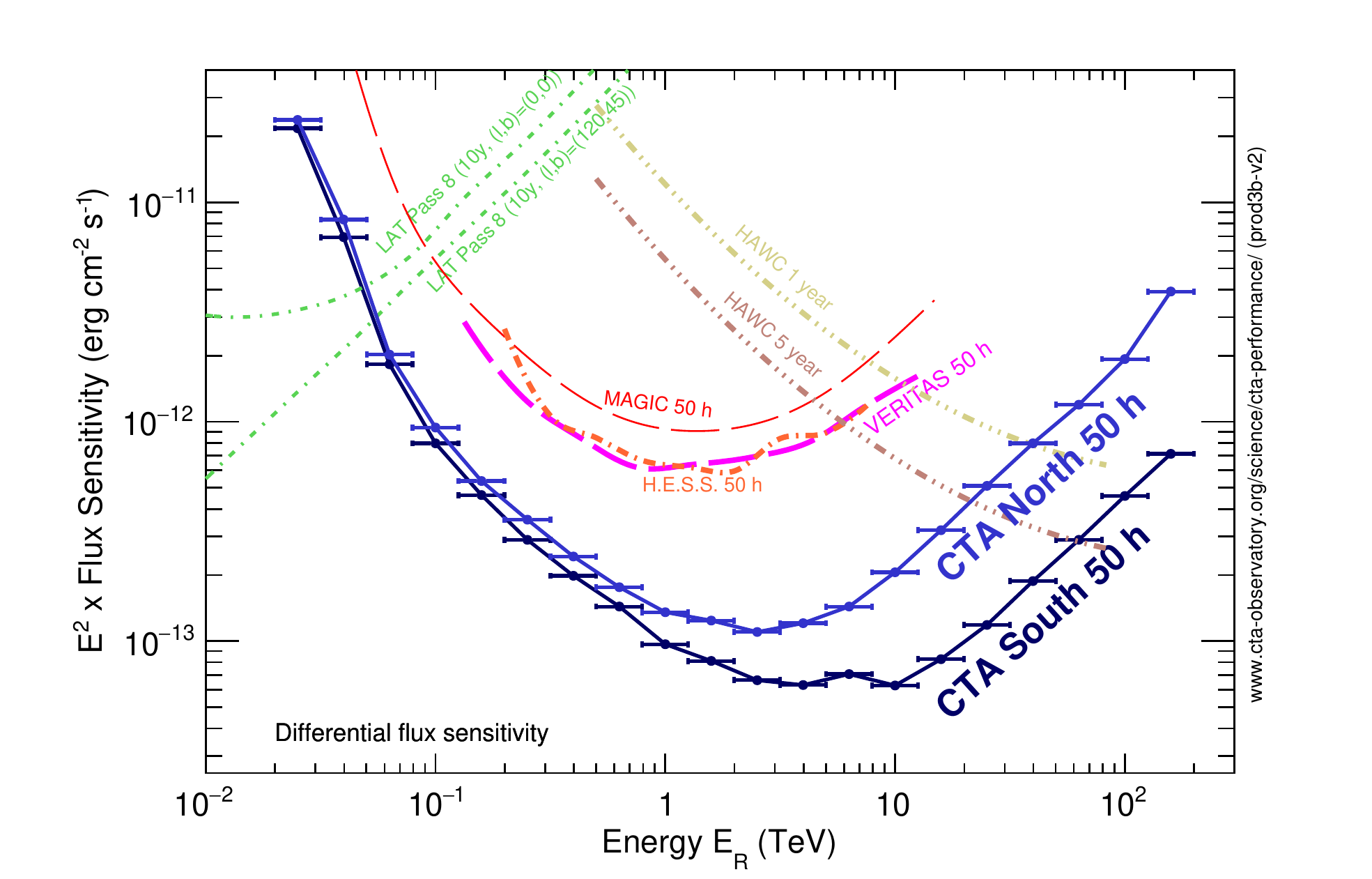}
\caption{\label{fig:diffsensOthers}
Differential energy flux sensitivities for CTA North and South calculated for 50 h of observation time in comparison with sensitivities of the Fermi LAT \cite{Latperf}, H.E.S.S. \cite{HESSperf}, MAGIC \cite{MAGICperf}, VERITAS \cite{VERITASperf}, and HAWC \cite{HAWCperf}.
The curves for Fermi-LAT and HAWC are scaled by a factor 1.2 relative those provided in the references, to account for the different energy binning. 
The curves shown allow only a rough comparison of the sensitivity of the different instruments, as the method of calculation and the criteria applied are not identical. 
In particular, the definition of the differential sensitivity for HAWC is rather different due to the lack of an accurate energy reconstruction for individual photons in the HAWC analysis.}
\end{figure}

The most important performance benchmark is the \textbf{differential sensitivity}, shown in Figures \ref{fig:diffsens}, \ref{fig:diffsensOff}, and \ref{fig:diffsensNSB} for point-like gamma-ray sources for the two CTA sites and for different observing conditions.
Differential sensitivity is defined as the minimum flux needed by CTA to obtain a 5-standard-deviation detection of a point-like source, calculated in non-overlapping logarithmic energy bins (five per decade). 
Besides the significant detection, we require at least ten detected gamma rays per energy bin, and a signal/background ratio of at least 1/20.  
The analysis cuts in each bin have been optimised to achieve the best flux sensitivity to point-like sources. 
The optimal cut values depend on the duration of the observation, therefore the instrument response functions are provided for three different observation times: 0.5, 5 and 50 hours.

Figure \ref{fig:diffsensOthers} compares the sensitivity of CTA with those of other major instruments in the field. 
The figure shows the significant improvement CTA will provide, especially in the energy range between 50 GeV and 25 TeV. For a comprehensive overview of how CTA will transform our understanding of the high-energy universe and which fundamental questions will be answered, see the CTA science book \cite{CTAScience}.

Instrument response functions for the observing conditions discussed in these proceedings can be downloaded from \cite{CTAPerf}.

\section*{Acknowledgments}

\noindent This work was conducted in the context of the CTA Consortium and CTA Observatory.

\noindent We gratefully acknowledge financial support from the agencies and organizations listed here: 
\url{http://www.cta-observatory.org/consortium_acknowledgments}.
We also would like to thank the computing centres that provided resources for the generation of the instrument response functions, see \cite{CTAPerf} for a full list of contributing institutions.


\begin{thebibliography}{99}
\bibitem{CTA} \url{www.cta-observatory.org}
\bibitem{Funk:2012} Funk, S. \& Hinton, J. (The CTA Consortium), 2013, Astroparticle Physics,  43, 348
\bibitem{CTAMC} Bernl\"{o}hr, K. et al. (The CTA Consortium), 2013, Astroparticle Physics, 43, 171
\bibitem{corsika} D. Heck et al., CORSIKA: a Monte Carlo code to simulate extensive air showers, 1998, Tech. Rep. FZKA 6019, Forschungszentrum Karlsruhe
\bibitem{Simtelarray} Bernl\"{o}hr, K., 2008, Astroparticle Physics, 30, 149
\bibitem{Eventdisplay} Maier, G. \& Holder, J., 2017, \pos{PoS(ICRC2017)747}. arXiv:1708.04048 
\bibitem{MARS} Moralejo, A., Gaug M., Carmona E. et al., 2009, Proceedings of the 31st International Cosmic Ray Conference, \L{}odz, arXiv:0907.0943
\bibitem{CTAsites} Hassan, T. et al., 2017, Astroparticle Physics 93, 76
\bibitem{CTAarrays} Acharyya, A. et al. (The CTA Consortium), 2019, Astroparticle Physics 111, 35
\bibitem{CTAScience}  The Cherenkov Telescope Array Consortium, Science with the Cherenkov Telescope Array
\bibitem{CTAPerf} \url{https://www.cta-observatory.org/science/cta-performance}
\bibitem{Latperf} \url{http://www.slac.stanford.edu/exp/glast/groups/canda/lat_Performance.htm}
\bibitem{HESSperf} Holler et al (The H.E.S.S. collaboration),  2015 Proceedings of the 34th ICRC (adapted)
\bibitem{MAGICperf} Aleksi\'{c}, J. et al. (The MAGIC Collaboration, 2016  Astroparticle Physics 72, 76
\bibitem{VERITASperf} \url{http://veritas.sao.arizona.edu/about-veritas-mainmenu-81/veritas-specifications-mainmenu-111}
\bibitem{HAWCperf} Abeysekara, A.U. et al. (The HAWC Collaboration), 2017, ApJ 843, 39
(World Scientific Publishing, 2019), ISBN 978-981-3270-08-4, arXiv: 1709.07997
\end{thebibliography}
\end{document}